\documentstyle[epsf,12pt]{article}
\newcommand{\insertplot}[1]{
\begin{center}\leavevmode\epsfysize=9.0cm \epsfbox{#1}\end{center}}

\begin{document}


\title{Strange hyperon and antihyperon production from
quark and string-rope matter}

\author{ P. Csizmadia$^{1}$,  P. L\'evai$^{1,2}$\footnote{Talk given at
the 4th Int. Conference on Strangeness in Quark Matter, July 20-24, 1998,
Padova, Italy, submitted to J. Phys. G.}, S.E. Vance$^{2}$, \\  
T.S. Bir\'o$^{1}$, M. Gyulassy$^{2}$, J. Zim\'anyi$^{1}$\\[2ex]
$^{1}$ KFKI Research Institute for Particle and Nuclear Physics, \\ 
\ \  P. O. Box 49, Budapest, 1525, Hungary \\[1ex]
$^{2}$ Department of Physics, Columbia University, \\
 New York, NY, 10027, USA
}

\date{18 September 1998}
\maketitle

\begin{abstract}
Hyperon and antihyperon production is investigated using two microscopical 
models: {\bf (1)} the fast hadronization of quark matter 
as given by the ALCOR model;  {\bf (2)} string formation and fragmentation
as in the HIJING/B model.  We calculate the particle numbers and momentum 
distributions for Pb+Pb collisions at CERN SPS energies in order to compare
the two models with each other and with the available 
experimental data.  We show that these two theoretical approaches
give similar yields for the hyperons, but strongly differ for antihyperons.  

\end{abstract}



\section{Introduction}
\label{sec1}
 
The quest to produce the Quark-Gluon Plasma (QGP) at the CERN SPS has reached
its final stage with the Pb+Pb run at $E_{beam}=158 \ {\rm AGeV}$. 
Since strangeness enhancement in heavy ion collisions relative to $pp$
collision has been predicted as a QGP
signal \cite{Bir1,Raf}, the measurement of strange and antistrange particles, 
especially the hyperons and the antihyperons, has received much attention
(see Refs.~\cite{NA49S98grab}-\cite{NA49S96}).
Several thermal models were invented to reproduce
earlier experimental data at the AGS and the SPS energies 
\cite{Raf2,Braun, Sollf1}.
However, only extended versions which introduce 
more parameters are capable of fitting the new 
experimental data which particularly include the hyperon and 
antihyperon yields~\cite{Raf3}.
Although these thermal models are capable of describing the gross 
yields and thermodynamic properties which are observed in the final state
of the collisions, they are unable to provide a description of the various
underlying microscopical mechanisms which may be involved in the initial
stages following the collision of the two nuclei.  Thus, to test for the 
formation of a QGP or any other possible initial phase, we choose two 
microscopical
models which describe the evolution of heavy ion collisions from two 
different viewpoints; a plasma vs. a hadronic scenario.

The first model assumes the formation of a deconfined 
state (namely the Constituent Quark Plasma (CQP) \cite{cqm1,cqm2}) 
and then describes its hadronization. 
The applied ALgebraic Coalescence Rehadronization (ALCOR) model 
\cite{ALCOR1,ALCOR2} and  MIcroscopical Coalescence Rehadronization 
(MICOR) model \cite{MICOR} (as well the
time-dependent Transchemistry Model \cite{cqm1,cqm2})
are all based on this assumption.

The second model begins with a  description
of hadron-hadron collisions using string phenomenology.
The strings are
then fragmented to produce the final observed hadron spectrum. 
The HIJING/B \cite{vance_hijb} event generator (an extension of 
HIJING \cite{hijing} event generator to include a baryon number
transport mechanism) is one version of this model.   
The Relativistic Quantum Molecular Dynamics (RQMD) model is another variant
of this model \cite{RQMD}.  
However, we note that HIJING and HIJING/B greatly differ from
RQMD in that they include minijets.

Although both of these models are able to reproduce pion, kaon and proton 
data, they differ in their predictions of the hyperons and the antihyperons. 
Thus the underlying assumptions of these models can 
be investigated by comparing their theoretical predictions for the 
hyperon and the antihyperon production with the experimental data. 

Before discussing our calculations, we argue that
secondary hadronic collisions negligibly effect the measured hadronic 
spectra at SPS energies. 
We note that secondary hadronic interactions are needed to  
describe the lower energy heavy ion collisions (at SIS and AGS) as seen in
the successful use of hadronic cascade models in reproducing data at these
energies. However, at $E_{beam} \approx$ 100-200 GeV/nucleon
we expect that the collisions involve the quarks of the nucleon. 
This introduces uncertain hadronic formation time with 
the production of a hadron from the intermediate quark matter 
or string/rope resonance.  This time is on the order of the 
life-time of the early highly excited system.  This idea is just the old
argument given by Pomeranchuk and Feinberg \cite{Pomer} who claimed
that particles only really materialize, not at the early collision
stage, but at low energy densities where they freeze-out and become
interactionless; the ``free separation" of particles.

The HIJING and HIJING/B models use this assumption,
since fragmentation occurs after the nuclei have finished colliding.
(We will also display the 'no-scattering' results of
RQMD from Ref.~\cite{RQMD}.)

The ALCOR and MICOR models assume that the hadronization of deconfined matter 
yields an excited hadronic gas which includes excited resonances,
namely the vector meson nonet and the baryon decuplet beyond the
lowest lying pseudoscalar nonet and baryon octet. The resonances
decay into the stable hadrons.  All of these models, except RQMD  
neglect final state interactions.  

Since the final hadron production 
of these models is directly related 
to the underlying microscopical mechanisms, we compare 
their predictions of hyperon and antihyperon formation 
with each other and with the available experimental data.

\section{The ALCOR and MICOR models}

In Pb+Pb collisions at the CERN SPS energies, it is assumed that  
the incoming nucleons disintegrate into a form of deconfined
matter. Originally the appearance of the perturbative 
QGP phase was expected. However, around the critical temperature,
$T_c \approx 150-200$  MeV,
perturbative QCD is unable to describe the characteristic
processes completely, 
thus non-perturbative or phenomenological descriptions
should be introduced for the deconfined matter.
The non-perturbative
descriptions can be motivated by lattice QCD calculations of
the equation of state near critical temperature $T_c$. Recent analysis
indicates
\cite{Lev1} that deconfined matter in its equilibrium
state can be described by massive quarks and gluons. 
These gluon-like quasi-particles are heavier than the quarks 
(thus their number is suppressed if the system is thermalized) 
and the effective quark masses are 
close to the constituent quark masses of the additive quark model 
\cite{RujGlas}.  We call this deconfined state the Constituent Quark 
Plasma (CQP). The CQP state is dominated by quarks and antiquarks, 
where some combine to form colorless hadron-like objects.
The presence of such prehadrons in the deconfined phase is confirmed
by lattice-QCD calculations of hadronic correlation functions 
\cite{Hadcorr}.   It is assumed that 
these prehadrons can escape from the strongly interacting region
to become mostly excited ``real'' hadrons.  

The ALCOR model describes this
microscopical hadronization process using quark coalescence
(for a detailed description see Refs.\cite{ALCOR1,ALCOR2}).
The coalescence rates depend on the temperature $T$, on the thermal quark
masses $M_q$ and $M_s$, and on the strength of the coalescence, which
can be characterized by an effective $\alpha_{eff}$ coupling
in a Coulomb-like quark-quark interaction potential. From
the point of view of hyperon and antihyperon production it is
important to note, that  the applied coalescence cross section is 
proportional to the mass of produced prehadron, 
$\sigma_{coal}^{i+j\rightarrow h} \propto m_h^2$ 
(see Ref.~\cite{ALCOR1}). Thus the production of heavier particles,
especially the hyperons (antihyperons) containing the heavy strange quark 
(antiquark) is relatively enhanced.
To describe the hadron formation completely, two more parameters
are needed: the number of newly produced light quarks 
$N_{q\overline{q}}=N_{u\overline{u}}=N_{d\overline{d}}$
and number of strange quarks $N_{s\overline{s}}$.  
The values of $N_{q\overline{q}}$ and $N_{s\overline{s}}$ are scaled by
$A^\beta$, where $\beta \approx 1$.
These input parameters, combined with some initial geometry
and a longitudinal Bjorken scaling are the basis
of the ALCOR model which calculates the total number of produced hadrons
for different heavy ion collisions.

In the ALCOR model the coalescence rates are determined from
multidimensional integrals which can be calculated
exactly for the case of Bjorken scaling.
To obtain the differential momentum spectra
of the produced hadrons, integrals need to be separated and
differential coalescence rates need to be evaluated numerically.
The introduction of numerical integrals allows us to consider
a more complicated flow pattern, in the present case
a constant transverse flow characterized by
$v_{tr}$, and a finite longitudinal extension characterized by
the maximal space-time rapidity $\eta_{max}$. 
A new version of the ALCOR model, the Microscopical Coalescence 
Rehadronization (MICOR) model \cite{MICOR}, includes this new  
flow pattern and calculate the differential momentum spectra.
Here, the microscopical quark coalescence produces off-shell prehadrons
and we assume that these prehadrons escape the deconfined region
conserving their velocity to become real on-shell hadrons.
(Because of the small deviation between prehadron and real hadron
masses, this assumption works well.)
After calculating the differential rates, the absolute norms of the
particle production are determined by a set of equations based
on quark and antiquark number conservation, similarly to ALCOR.
The MICOR model is able to determine the differential hadron spectra 
which arises from the hadronization of the CQP state.
As in ALCOR, final state hadronic rescattering is neglected.

The advantage of using the ALCOR and MICOR models is
that the phase transition can be followed microscopically.  In this way a
phenomenological quark-quark interaction determines the formation 
of a chemically out-of equilibrium hadronic gas.

\section{The HIJING/B model}
As has been summarized elsewhere \cite{hijing}, 
the HIJING Monte Carlo event generator
models hadronic interactions by combining low $p_T$ multistring 
phenomenology with the perturbative QCD (pQCD) processes.
The HIJING/B~\cite{vance_hijb} event generator extends the HIJING model
by introducing a novel, non-perturbative gluon junction mechanism 
in order to study baryon number transport in heavy-ion collisions.
As shown below, this novel non-perturbative junction mechanism strongly
influences hyperon production. 

This gluonic mechanism is motivated from the non-perturbative gluon field
configuration (the baryon junction) that appears
when writing the simplest gauge
invariant operator for the baryon in ${\rm SU}_c(3)$;
\begin{eqnarray}
B &=& \epsilon^{ijk} 
\left [ P \exp \left (ig \int_{x_1}^{x_J} dx^{\mu} A_{\mu} 
\right ) q(x_1) \right ]_i \nonumber \\
&\times& \left [P \exp \left (ig \int_{x_2}^{x_J} dx^{\mu}
A_{\mu} \right ) q(x_2) \right ]_j 
\left [P \exp \left (ig \int_{x_3}^{x_J} dx^{\mu} A_{\mu}
\right ) q(x_3) \right ]_k.
\end{eqnarray}
Here, the baryon junction is the vertex at $x_J$ where the three gluon
Wilson lines link the three valence quarks
to form the gauge invariant non-local operator.
In a highly excited baryonic state, the Wilson lines represent
color flux tubes.   When these strings fragment via $q\bar{q}$ production,
the resulting baryon will be composed of the three sea quarks
which are linked to the junction while the original valence quarks will
emerge as constituents of three leading mesons.  Being a gluonic
configuration, it was proposed~\cite{khar_bj96} that the junction
could be easily transported into the mid-rapidity region
in hadronic interactions.

In Regge phenomenology, allowing for the possibility of baryon junctions
in scattering is taken into account by introducing additional Regge
trajectories ($M^J$).    As was shown by Kharzeev~\cite{khar_bj96},
the exchange of these trajectories in $p+p$ interactions leads to events
where one of the baryon junctions is stopped in the mid-rapidity 
region.  Once the baryon junction is stopped, the three valence quarks 
continue forming three beam jets.   These events have a different energy, 
rapidity and multiplicity dependence compared to 
the normal Pomeron two string event.
In particular, the produced baryon has a $\cosh(y/2)$ rapidity dependence
and a $1/\sqrt[4]{s}$  energy dependence. 
In addition, the new string configuration leads to a 
multiplicity enhancement of a factor of 5/4 (counting the two + three  
jets as compared with the two + two jets).   
Since the baryon which is resolved around the 
junction in the mid-rapidity region is composed of three sea quarks, 
hyperon production 
and the  $<p_t^2>$ of the final baryon are enhanced by factors of 3.   
The enhancement factor of 3 of the strangeness content
of the baryon allows for the unique possibility of
producing $\Omega^-$ ($S=-3$) baryons.

\newpage

In HIJING/B~\cite{vance_hijb}, the baryon junction mechanism is implemented
using a ``Y'' string configuration for the excited baryon with an
effective cross section of $\sim 9$ mb.    The above junction dynamics
does not provide a mechanism for antibaryons production and further
studies which address baryon pair production are underway.  In their present
version, neither HIJING nor HIJING/B include final state interactions. Thus,
in one variant of HIJING/B we simulate multiple final state interactions 
using the concept of ``ropes'' \cite{Biro_rope};  
increasing the energy per unit length of the strings from
1.0 GeV/fm  to 1.45 GeV/fm in the heavy nuclear collisions.
Increasing the effective string tension increases 
the strangeness production and the $<p_T>$.

\section{Results on PbPb collision}

In the ALCOR model, we recalculate the hadron production as compared
with earlier publications \cite{ALCOR1,ALCOR2}.  In this calculation, 
we considered the feeding of the detected $\Lambda$s from
$\Xi$ and $\Omega$ decay channels and the undistinguishable $\Sigma^0$. 
We also introduce the 
$\Lambda^0-$ like $\equiv\Lambda^0+\Sigma^0+\Xi^-+\Xi^0+\Omega^-$ species,
which was not done earlier.
This number will be compared to the experimentally measured 
number of ``$\Lambda$"-particles, as was done
similarly with ``$\overline{\Lambda^0}$-like" particles.
In this case the
input parameters for newly produced quark pairs are $N_{q\overline q}=391$ 
and $N_{s\overline s}=172$ (which means $g_s = N_{s\overline s}/
(N_{u\overline u} + N_{d\overline d}) = 0.22$)
and for the effective quark-quark interaction,
$\alpha_{eff}=0.97$. The results are displayed in Table 1.

The particle yields were also calculated in HIJING/B with ropes
where the energy per unit length of the string was increased 
to $\kappa=1.45$ GeV/fm.  The total numbers are shown in Table 1 
along with the RQMD results \cite{RQMD}.

When comparing \ the \ two \ calculations, ALCOR and HIJING/B predict 
approximately the same numbers of mesons, baryons and 
non-strange antibaryons; approximately reproducing the experimental data. 
However, significant differences can be seen in the predictions of 
strange antibaryons and the $\Omega$.   These differences
are directly related to the assumed microscopical hadronization mechanisms 
and thus reveal the applicability of the different assumptions.
These differences can be
investigated via particle ratios, which are displayed in Table 2 and 3.
In Table 2 and in Table 3, the experimental results obtained from NA49 
Collaboration and the WA97 Collaboration are shown, respectively.
We note that these measured ratios are valid in certain rapidity windows, 
and thus only provide an approximate comparison.

We also investigate the energy dependence of our results.
Particle abundances were calculated at 100, 200 and 300 AGeV for Pb+Pb
collisions using HIJING/B with ropes 
and the ALCOR model.  In the ALCOR model we kept
the $g_s$ and $\alpha_{eff}$ parameters at their value obtained
by fitting the data at 158 AGeV, and only changed the number of newly 
produced light quarks, $N_{q{\overline q}}=354,442,496$, respectively, to
reproduce the $h^-$ number as obtained in HIJING/B.  The 
results of the energy dependence of strange particle ratios from these
calculations are shown in Figure 1.  In addition, we show
the experimental results from the NA49 and WA97 experiments.
Although both models reveal a slight energy dependence, only 
the ALCOR model was able to consistently reproduce the observed values. 

\begin{center}
\begin{tabular}{||c||c||c|c|c||}
\hline
{\bf Pb+Pb} & {\bf NA49}
& {\bf ALCOR } & {\bf HIJING/B} & {\bf RQMD } \\
\hline
\hline
 $h^{-}$  & $680^a$ 
&679.8    &700.8    &829.2 \\
\hline
\hline
 $\pi^+$  &  
&590.6    &608.7    &692.9  \\
\hline
 $\pi^0$  &  
&605.9    &622.8    &724.9  \\
\hline
 $\pi^-$  &  
&622.0    &636.9    &728.8  \\
\hline
 $K^+$    & $76^*$  
&\ 78.06  &\ 79.4   &\ 79.0   \\
\hline
 $K^0$    &  
&\ 78.06  &\ 79.4   &\ 79.0  \\
\hline
 ${\overline K}^0$    &  
&\ 34.66  &\ 35.2   &\ 50.4  \\
\hline
 $K^-$    & $\{32\}^b $  
&\ 34.66  &\ 35.2   &\ 50.4   \\
\hline
\hline
 $p^+$    &  
&153.2    &152.3    &199.7  \\
\hline
 $n^0$    &  
&170.5    &165.2    &217.6  \\
\hline
 $\Sigma^+$    &  
&\ \ 9.16 &\ 12.3   &\ 12.9  \\
\hline
 $\Sigma^0$    &  
&\ \ 9.76 &\ 11.8   &\ 13.1     \\
\hline
 $\Sigma^-$    &  
&\ 10.39  &\ 12.4   &\ 13.3    \\
\hline
 $\Lambda^0$   &  
&\ 48.85  &\ 40.2   & \ 35.3  \\
\hline
 $\Xi^0$  &  
&\ \ 4.89 &\ \ 4.90 & \ \ 4.2   \\
\hline
 $\Xi^-$  & $7.23 \pm 0.88^d$ 
&\ \ 4.93 &\ \ 4.93 & \ \ 4.2   \\
\hline
 $\Omega^{-}$  &  
&\ \ 0.62 &\ \ 0.43 &   \\
\hline
\hline
 ${\overline p}^-$   &  
&\ \ 6.24 &\ 9.67   &\ 27.9    \\
\hline
 ${\overline n}^0$   &  
&\ \ 6.24 &\ 9.62   &\ 27.9   \\
\hline
 ${\overline \Sigma}^-$   &  
&\ \ 0.91 &\ \ 1.33 &\ \ 4.6   \\
\hline
 ${\overline \Sigma}^0$   &  
&\ \ 0.91 &\ \ 1.19 &\ \ 4.6    \\
\hline
 ${\overline \Sigma}^+$   &  
&\ \ 0.91 &\ \ 1.17 &\ \ 4.6    \\
\hline
 ${\overline \Lambda}^0$   & 
&\ \ 4.59 &\ \ 2.84 &\ 10.7  \\
\hline
 ${\overline \Xi}^0$   &  
&\ \ 1.12 &\ \ 0.44 &\ \ 2.0    \\
\hline
 ${\overline \Xi}^+$   & $\{1.6 \pm 0.2 \}^d$  
&\ \ 1.12 &\ \ 0.44 &\ \ 2.0    \\
\hline
 ${\overline \Omega}^{+}$   &  
&\ \ 0.35 &\ \ 0.012 &   \\
\hline
\hline
 $K^0_{S}$   &$ \{54\}^{b,c}$  
&\ 56.36  &\ 57.3   & \ 63.5  \\
\hline
$p^+-{\overline p}^-$ &$\{145\}^a$ 
&147.03   &142.6   &171.8 \\
\hline
 $\Lambda^0$-like &$\{49\pm 6\}^d$  
&\ 69.07  & \ 62.31 &\  56.8    \\
\hline
 ${\overline \Lambda}^0$-like &$\{8.8\pm 1.0\}^d$  
&\ 8.12   & \ 4.92  &\  19.3   \\
\hline
\hline
\end{tabular}
\end{center}
\vskip 0.5truecm
\noindent {\bf Table 1:}
Total hadron multiplicities for $Pb+Pb$ collision
at bombarding energy 158 GeV/nucleon. The displayed
experimental results are obtained or estimated $(\{ \})$ from the
publications of NA49 Collaboration:
${}^a$ is from \cite{NA49QM96};
${}^b$ is from \cite{NA49S97}; ${}^c$ is from
\cite{NA49S96}; 
${}^d$ is from \cite{NA49S98grab};
${}^*$ is estimated from $\{K^-\}$ and
$\{K^0_S\}$. 
Theoretical results are calculated in
the ALCOR and HIJING/B models and it
was included the results of RQMD model 
("ropes + no re-scattering" version) \cite{RQMD}. 
We introduced
$\Lambda^0-$ like $\equiv\Lambda^0+\Sigma^0+\Xi^-+\Xi^0+\Omega^-$.
\bigskip

\begin{center}
\begin{tabular}{||l||c|c|c|c||}
\hline
{\bf Pb+Pb} &  {\bf NA49}  & {\bf ALCOR} &{\bf HIJING/B} & {\bf RQMD} \\
\hline
 ${\overline {\Xi}}^+/\Xi^-$    
 &$0.232 \pm 0.033$ &0.227 & 0.089 & 0.476   \\
\hline
 ${\overline {\Xi}}^+/{\overline {\Lambda}}$                
 &$0.188 \pm 0.039$ &0.138& 0.089 & 0.103  \\
\hline
 $\Xi^-/\Lambda$ 
 &$0.148 \pm 0.011$ &0.071& 0.078 &0.074 \\
\hline
\hline
\end{tabular}
\end{center}
\noindent{\bf Table 2.}
Strange baryon and antibaryon ratios measured by
NA49 Coll. \cite{NA49S98grab}
for $Pb+Pb$ collision at 158 GeV/nucl bombarding energy 
in the momentum
region $3.1 \geq y \geq 4.1$, $p_T > 0$.
We calculated the ratios of total numbers from Table 1.
Here $\Lambda$($\overline \Lambda$) means
``$\Lambda^0$-like" (``${\overline \Lambda}^0$-like") species.
\bigskip

\begin{center}
\begin{tabular}{||l||c|c|c|c||}
\hline
\hline
{\bf Pb+Pb} &  {\bf WA97}  & {\bf ALCOR} &{\bf HIJING/B} & {\bf RQMD}  \\
\hline
\hline
 ${\overline {\Lambda}}/\Lambda$    
 &$0.128 \pm 0.012$ &0.117  & 0.079  & 0.339  \\
\hline
 ${\overline {\Xi}}^+/\Xi^-$    
 &$0.266 \pm 0.028$ &0.227  & 0.089  & 0.476  \\
\hline
 ${\overline {\Omega}}^+/\Omega^-$                
 &$0.46  \pm 0.15$  &0.564  & 0.028  & ---    \\
\hline
 $\Xi^-/\Lambda$                
 &$0.093 \pm 0.007$ &0.071  & 0.079  & 0.073  \\
\hline
 ${\overline {\Xi}}^+/{\overline {\Lambda}}$                
 &$0.195 \pm 0.023$ &0.138  & 0.089  & 0.103  \\
\hline
 $\Omega^-/\Xi^-$ 
 &$0.195 \pm 0.028$ &0.125  & 0.087  & ---    \\
\hline
\hline
\end{tabular}
\end{center}
\noindent{\bf Table 3.}
The ratios of strange baryons and antibaryons,  measured by the
WA97 Collaboration \cite{WA97QM97} 
for $Pb+Pb$ collision at 158 GeV/nucl bombarding energy at $p_T > 0 \ GeV$.
We calculated the ratios of total \ numbers \  from  Table 1.
Here $\Lambda$($\overline \Lambda$) means
``$\Lambda^0$-like" (``${\overline \Lambda}^0$-like") species.
\bigskip

\insertplot{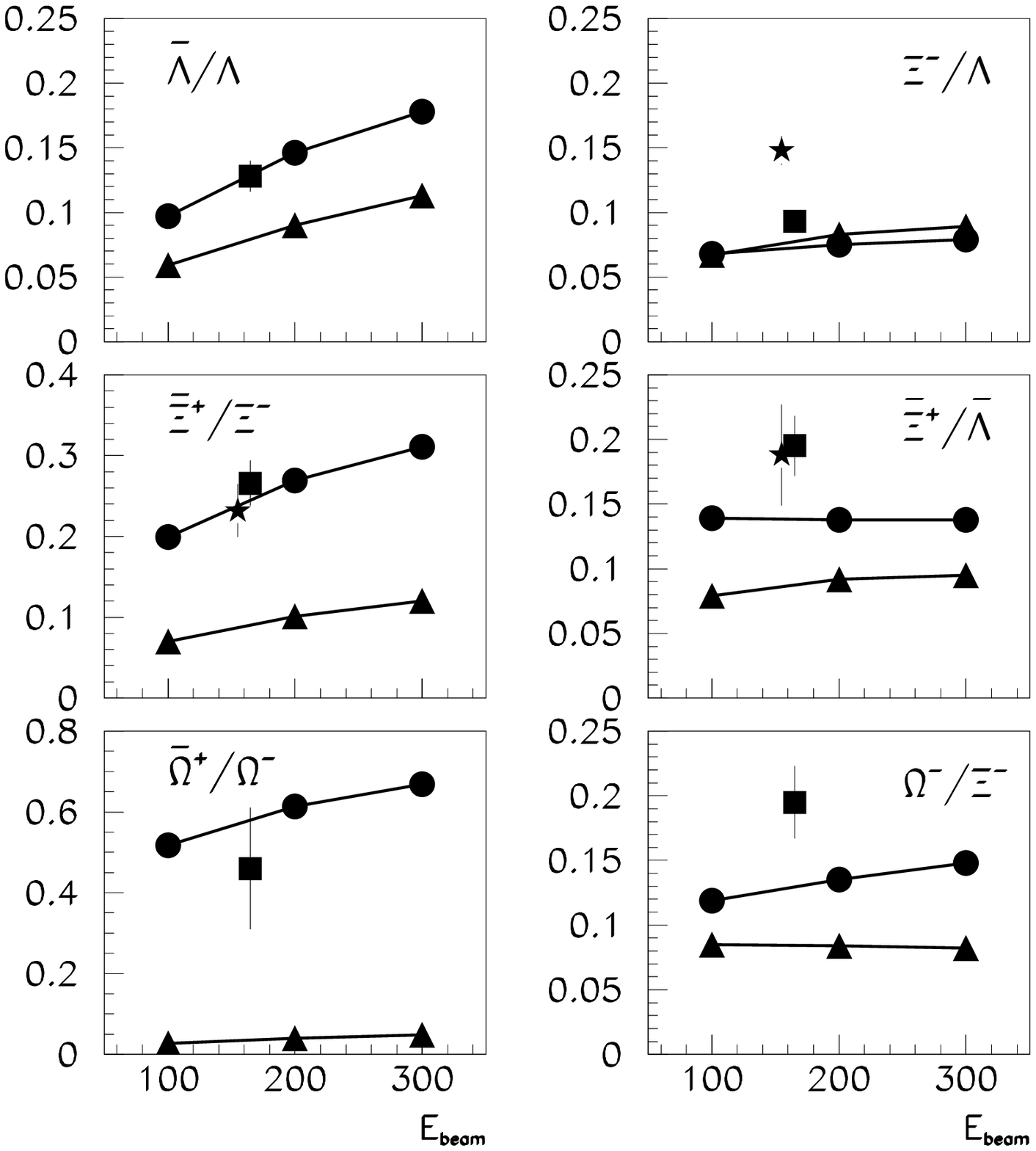}
 \begin{center}
  \begin{minipage}[t]{13.054cm}
  { {\bf Figure 1.}
  The measured ratios of baryons and antibaryons
  in the PbPb collision at 158 AGeV energy from
  the NA49 Coll. (stars) \cite{NA49S98grab,NA49QM96}
  and the WA97 Coll. (squares) \cite{WA97QM97}. The
  theoretical results from ALCOR (dots) and HIJING/B (triangles)
  are displayed at 100, 200 and 300 AGeV energy.
   }
  \end{minipage}
 \end{center}

The transverse momentum distributions of the produced hadrons in 
the mid-rapidity region were also calculated using the MICOR model. 
For this calculation, two initial conditions were considered. 
Since the ALCOR model is not very sensitive to the
initial temperature $T_{CQP}$, 
in the range $150 \ {\rm MeV} < T_{CQP} < 180 \ {\rm MeV}$, 
two set of parameters 
were used in the MICOR calculation: 
I. $T_{CQP}=150 \ {\rm MeV}$, $v_{tr}=0.67$;
II. $T_{CQP}=175 \ {\rm MeV}$, $v_{tr}=0.57$. 
In case I, the pion and proton spectra is reproduced at midrapidity.
In case II, good agreement was found for the heavier hyperons.

\insertplot{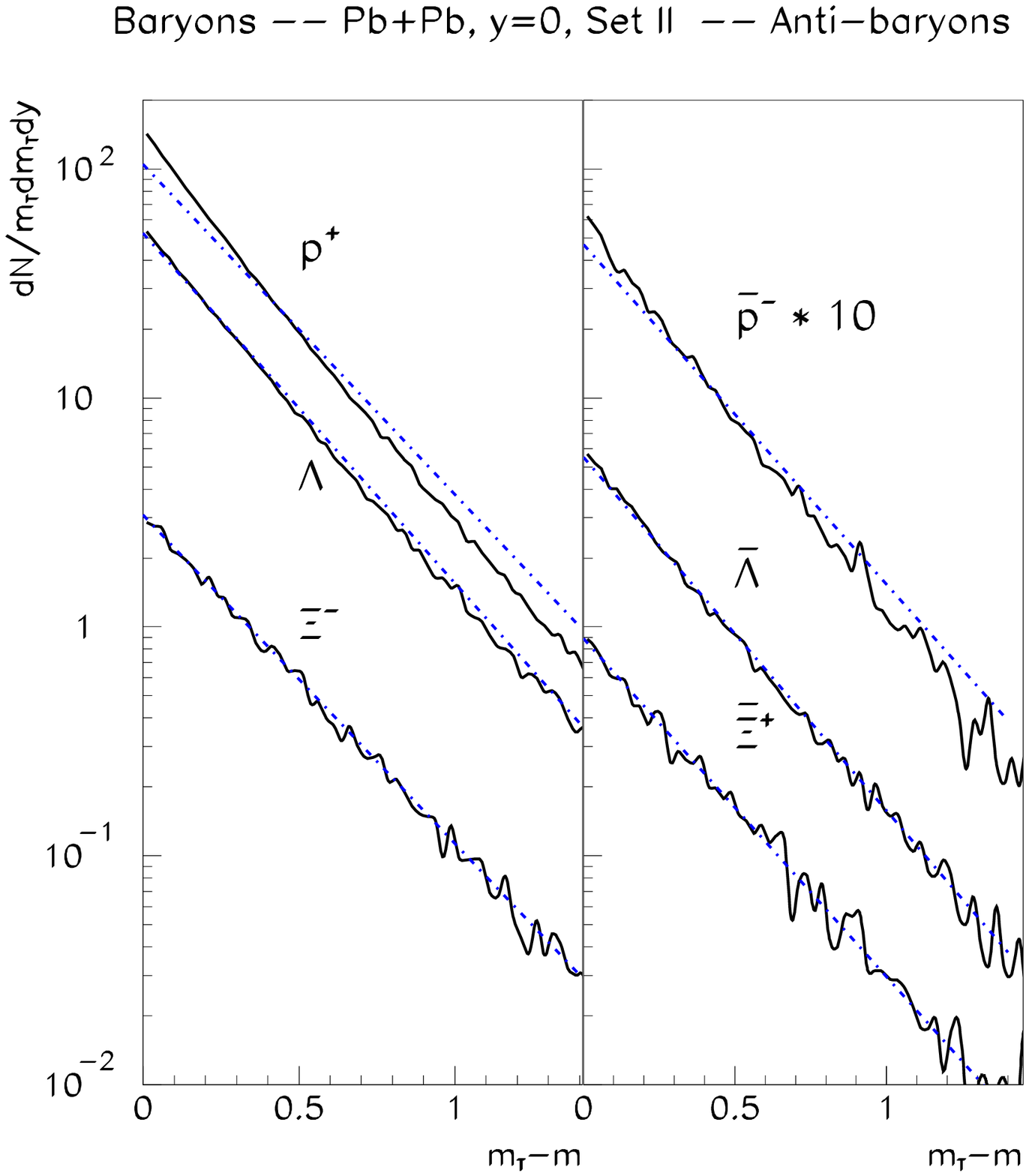}
 \begin{center}
  \begin{minipage}[t]{13.054cm}
  { {\bf Figure 2.}
  The transverse momentum distribution
  of proton, $\Lambda$, $\Xi^-$
  and their antiparticles in Pb+Pb collision in midrapidity,
  produced by the MICOR model with initial condition set II.
  The full curves are the results of calculation and the dashed lines
  indicate experimental slopes measured by NA49 Coll. 
  \cite{NA49S98grab}.
  }
  \end{minipage}
 \end{center}

In the MICOR model, the non-thermal momentum distributions 
of the emerging excited hadrons can be calculated.  These distributions
can be characterized by a non-Boltzmann distribution,
$f\propto(m/E)^\delta \cdot \exp(-E/T)$, where $\delta \approx 2$.
What is remarkable is that after resonance decay the obtained hadronic spectra
can be very well fitted by pure exponential functions \cite{MICOR}, 
with the effective slopes closely matching the experimental
values.  In Figure 2, we display the resulting momentum spectra 
using the values from case II for the proton, $\Lambda$, $\Xi^-$ and 
their antiparticles.
Here, the solid lines show the MICOR results, which can be
parametrized by the exponential fit:

\begin{equation}
\frac{dN}{m_T dm_T dy} = C \cdot e^{-{m_T/T_{eff}}} \ .
\end{equation}

These results can be compared to the dashed lines which 
represent the slopes as 
measured by the NA49 Collaboration \cite{NA49S98grab,NA49S98mar}.
Here, we show that after two completely non-equilibrium steps
(quark coalescence and decay of excited hadrons) the final hadron
spectra is approximately exponential.  
Our result in Figure 2 demonstrates that
non-equilibrium processes are able to produce  particle spectra 
which can also be interpreted as thermal.
We note that baryons and antibaryons have approximately the same slope,
as has been observed.

In Figure 3, we display the effective slopes $T_{eff}$
as calculated using the MICOR model (using cases I and II)
and as measured by the NA49
\cite{NA49S98grab,NA49S98mar}, NA44 {\cite{NA44_97} 
and WA97 Collaborations \cite{WA97QM97,WA97S97}. 
The possible interpretation of our MICOR results is the following:
parameter sets I and II are so close to
each other that a mutual hadronization temperature
and transverse flow inbetween (namely $T_{CQP}\approx 170 MeV$ and
$v_{tr}\approx 0.6$) could be assumed for the produced
deconfined state as 
the dotted line shows in Figure 3 (see Ref.~\cite{NA49S98grab}).
The deviation of the recent experimental data from this line
indicates that
hyperons (antihyperons) which contain mostly strange (antistrange) quarks
escaped the light-quark dominated CQP state at a
slightly earlier time,
when the temperature is higher and the transverse flow is smaller.
Particles containing mostly light quarks are formed in the later
stage of the expanding CQP matter and thus their spectra indicate
a lower temperature but higher transverse flow upon hadronization.
We note that there is a similar, 
interacting hadronic explanation 
\cite{Sorge98} which was obtained from an RQMD-based analysis which assumes
secondary interactions with different strengths for strange and
non-strange particles.
We hope that final data on $\Omega^-$ and 
${\overline \Omega}^+$ production will help us to
make our conclusion more definite.

\insertplot{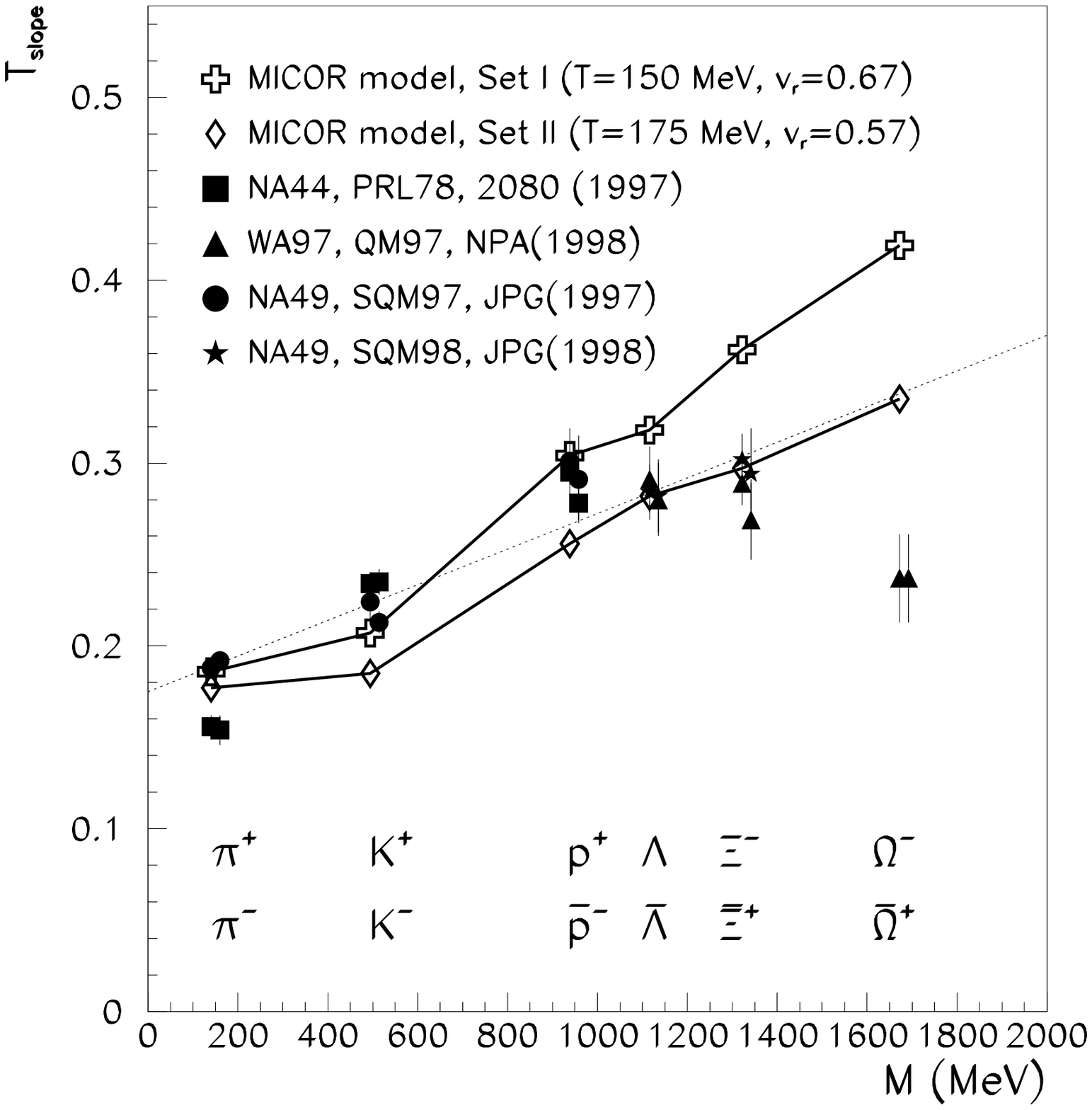}
 \begin{center}
  \begin{minipage}[t]{13.054cm}
  { {\bf Figure 3.}
  The effective slope parameters, $T_{eff}$, in the midrapidity region
  for Pb+Pb collision at 158 AGeV, obtained from the MICOR
  model and from the different experiments.
  The dotted line indicates the  
  appearance of a common transverse
  flow in the hadronic phase, see Ref.~\cite{NA49S98grab}.}
  \end{minipage}
 \end{center}

\section{Conclusion}
We investigate hyperon and antihyperon production in Pb+Pb collision at 
CERN SPS energies using two different scenarios. The formation of
a Constituent Quark Plasma and the modeling of its hadronization as 
in the ALCOR and MICOR models was used to calculate the particle numbers
and transverse momentum distributions. 
String formation and fragmentation as in the HIJING/B with ropes model 
was also used to calculate particle yields. 
In the case of meson, baryon and non-strange antibaryon yields, both 
models predict similar particle numbers.  However, only the CQP scenario 
could reproduce the antihyperon data.   This can be attributed to
the coalescence of heavier strange quarks which is favoured in the 
ALCOR model.  The inability of HIJING/B to reproduce the antihyperon 
data reveals the need for 
a new hadronic mechanisms (similar to the baryon junction) 
for antihyperon production.

In addition to reproducing the antihyperon yields, 
the results of the CQP-based MICOR model point to a scenario where 
a mutual transverse flow was developed in the deconfined phase 
and that hadronization 
occured when $T_{CQP}\approx 170$ MeV and 
$v_{tr}\approx 0.6$.
Here, hyperons and antihyperons may escape
a little bit earlier from the deconfined region, retaining a
slightly larger temperature and a smaller collective transverse flow.
The other hadrons are then formed shortly thereafter.


\section*{Acknowledgments}
One of us (P.L.) thanks for M. Morando for the kind hospitality 
at Padova and for F. Grabler and S. Margetis
to make available their contribution before submission.
One of us (M.G.) is greatful to the KFKI RMKI for its hospitality
during the course of this work.
This work was supported by the Hungarian Science Fund (OTKA) grants
T024094 and T025579 and
by the US-Hungarian Science and Technology Joint Fund No. 652/1998.

\end{document}